# Spin-orbit torques for current parallel and perpendicular to a domain wall

Tomek Schulz[1], Oscar Alejos[2], Eduardo Martinez[3], Kjetil M. D. Hals[4], Karin Garcia[5], Kyujoon Lee[1], Roberto Lo Conte[1,6], Gurucharan V. Karnad[1], Simone Moretti[3], Berthold Ocker[7], Dafiné Ravelosona[5], Arne Brataas[8], Mathias Kläui[1,6].

[1] Institut für Physik, Johannes Gutenberg-Universität Mainz, Staudinger Weg 7, 55128 Mainz, Germany

[2] Depatamento de Electricidad y Electromagnetismo, Universidad de Valladolid, Paseo de Belen, 7, E-47011, Valladolid, Spain

[3] Departamento Fisica Aplicada, Universidad de Salamanca, plaza de los Caidos s/n E-38008, Salamanca, Spain

[4] Niels Bohr International Academy and the Center for Quantum Devices, Niels Bohr Institute, University of Copenhagen, 2100 Copenhagen, Denmark

[5] Institut d'Electronique Fondamentale, UMR CNRS 8622, Université Paris Sud, 91405 Orsay Cedex, France

[6] Graduate School of Excellence "Materials Science in Mainz" (MAINZ), Staudinger Weg 9, 55128 Mainz, Germany

[7] Singulus Technologies AG, 63796 Kahl am Main, Germany

[8] Department of Physics, Norwegian University of Science and Technology, NO-7491, Trondheim, Norway

## Abstract

**We report field- and current-induced domain wall (DW) depinning experiments in Ta\Co$_{20}$Fe$_{60}$B$_{20}$\MgO nanowires through a Hall cross geometry. While purely field-induced depinning shows no angular dependence on in-plane fields, the effect of the current depends crucially on the internal DW structure, which we manipulate by an external magnetic in-plane field. We show for the first time depinning measurements for a current sent parallel to the DW and compare its depinning efficiency with the conventional case of current flowing perpendicularly to the DW. We find that the maximum efficiency is similar for both current directions within the error bars, which is in line with a dominating damping-like spin-orbit torque (SOT) and indicates that no large additional torques arise for currents parallel to the DW. Finally, we find a varying dependence of the maximum depinning efficiency angle for different DWs and pinning levels. This emphasizes the importance of our full angular scans compared to previously used measurements for just two field directions (parallel and perpendicular to the DW) and shows the sensitivity of the spin – orbit torque to the precise DW structure and pinning sites.**

# Manuscript

The development of novel magnetic device concepts is of major commercial as well as scientific interest to further improve today's state of the art low power information technology. Since established concepts such as tape drives, hard disk drives[1] and random access memory (RAM) have all their particular advantages, but also conceptual limitations, many novel non-volatile technologies have appeared in the last decade, e.g. NAND, FLASH, Phase change, resistive RAM or MRAM[2, 3]. Particularly promising memory candidates based on magnetic domain wall (DW) motion include the racetrack-memory device[4] and the DW MRAM device[5], based on current-induced motion of magnetic DWs in magnetic nanowires. The concept is particularly promising when using the spin-transfer torque (STT) or the even more efficient spin-orbit torques (SOTs)[6-10]. With a three terminal geometry, shifting, reading and writing currents can be decoupled, which circumvents the problem of tunnel barrier breakdown due to high current densities[11].

The current-induced motion of magnetic DWs in nanowires is usually modeled by the phenomenological Landau-Lifshitz-Gilbert (LLG) equation with STT terms to capture the interaction between spin-polarized conduction electrons and magnetic textures[12-14]. However, in material systems consisting of a heavy metal (HM)/ferromagnet (FM)/ Oxide (Ox) trilayer, ultrafast DW motion was observed in the direction against the electron flow[15, 16], which cannot be explained by the conventional STT.

Novel theoretical approaches for current-induced DW motion are taking into account additionally the influence of SOTs. These arise from the current flowing through the HM layer, in combination with a Dzyaloshinkii-Moriya interaction (DMI), which is present when the structural inversion symmetry is broken at the interfaces of a multilayer system[17].

Both the spin-Hall-effect (SHE)[18-20] and the inverse spin galvanic effect (also termed Rashba-Edelstein effect)[16,21] are considered as possible mechanisms generating the SOTs. They lead respectively to a damping-like and a field-like torque in a modified LLG equation and have been used to explain recent experimental results[6-10].

Following this model, the LLG equation (including the SOTs) predicts a maximum SOT efficiency for a current passing perpendicular to a DW if the wall has a Néel type spin structure (NW). Furthermore, by analyzing the SOTs, one sees that (in contrast to STT) even a current flowing parallel to the wall can displace the DW. In this case, the highest SOT efficiency is expected for DWs with a Bloch wall (BW) structure. This is due to the fact that the damping-like torque always acts on the magnetization component parallel to the current flow.

In a recent theoretical work, where a generalized symmetry-based approach was used to derive all possible contributions to the local torque-distribution-function $\tau(r,t) = m \times H_c[m, \nabla m, \mathcal{J}]$, the possibility of novel relativistic torques on the DW were predicted. Importantly, some of the predicted torques are shown to produce BW motion for currents applied parallel to the wall[22], but these predictions have not been tested, yet.

In this paper, we determine the strengths of the current induced torques on a DW for current flowing parallel and perpendicular to the DW. We study the depinning of a domain wall in a Hall cross (HC) as a function of various applied field vectors $\vec{H}$ and determine the efficiency for different DW structures using the current-field-equivalence-method[23-25]. Comparison between the torque strengths for different current and field directions allows us to ascertain the acting torques and compare them for current flowing parallel and perpendicular to the wall.

For the experiments, we use Si\SiOx\Ta(5nm)\Co$_{20}$Fe$_{60}$B$_{20}$(1nm)\MgO(2nm)\Ta(5nm) nanostructures exhibiting perpendicular magnetic anisotropy (PMA).

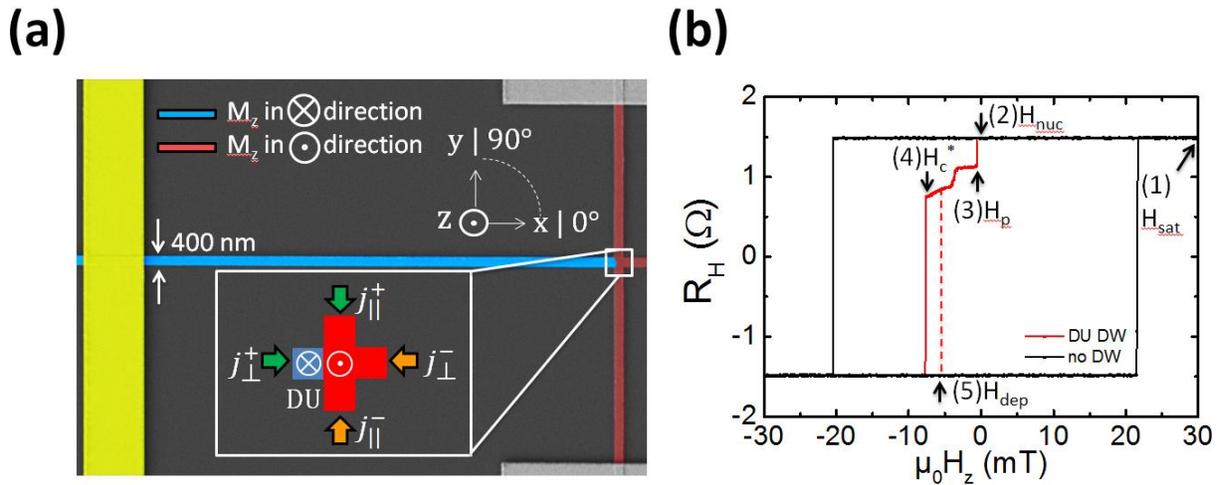

*Figure 1. (a): SEM image of the fabricated nanostructure consisting of a 400 nm wide nanowire with a Hall bar and a 1.4 µm wide Ti(5 nm)\Au(100 nm) Oersted field line*

*perpendicular and on top of the nanowire. The inset indicates the directions of the injected currents and the geometry of a down-up (DU) domain wall at the Hall cross. (b) shows the extraordinary Hall effect signal at the Hall cross (HC) for the case where: (i) no DW has been nucleated artificially in the nanowire (black line); (ii) a DU DW (red curve) has been nucleated by an Oersted field and depinned through the HC purely by using an applied negative out-of-plane field $H_c$\* (solid red line, 4); (iii) a DU DW has been depinned by an applied current density combined with a lower negative out-of-plane field $H_{dep}$ (dotted red line, 5) compared to the field needed to reverse the Hall cross without a nucleated domain wall (dotted black line)*

The thin film was deposited by sputtering using a Singulus TIMARIS/ROTARIS machine. The sample was annealed for 2 h at 300 °C in vacuum to generate a large PMA. Using SQUID measurements, we determined the in-plane magnetization saturation field to be $\mu_0 H_{sat} = 400\ mT$ and a saturation magnetization of $M_s = 1.1 \times 10^6\ A/m$. From these values, we obtained an effective anisotropy $K_{eff} = \mu_0 H_{sat} M_s / 2 = 2.2 \times 10^5\ J/m^3$. By using electron-beam lithography and argon-ion milling, the sample was patterned into a 400 nm wide nanowire with a Hall bar and a gold Oersted line on top of the nanowire. Fig 1(a) shows a scanning electron microscopy (SEM) image of the fabricated nanostructure.

For the experiments, we follow a measurement protocol whereby the magnetization at the position of the Hall bar is monitored after each field or current injection step by measuring the extraordinary Hall effect (Fig. 1b). For a Down-Up(DU)-DW (red-curve in Fig. 1b), we first saturate the nanowire towards the positive out-of-plane direction (indicated by point (1) in Fig. 1(b)). Then the out-of-plane field is relaxed to 0 mT (point (2)) and a 10 ns long current pulse of I≈100 mA is sent through the Au Oersted line, which is perpendicular to the magnetic nanowire. The Oersted field of the current pulse is strong enough to reverse locally the magnetization direction around the Oe line towards the -z-direction, thereby nucleating a new domain indicated by the blue color in Fig. 1(a).

After that an out-of-plane field is applied in the -z-direction to expand the nucleated magnetic domain pointing into the -z-direction, whereby the DU wall is shifted inside the Hall cross for fields of $|\mu_0 H_p| \approx 1\ mT$ (propagation field of the DW in the nanowire). Once the DW reaches the Hall cross, the Hall signal jumps to an intermediate value between the up and down saturation level (point (3) in Fig. 1(b)). This is our starting configuration, which allows us to study the DW depinning for currents parallel or perpendicular to the DW as a function of fields and currents.

To determine the strength of the torques, we measure the dependence of the depinning fields $H_{dep}$ of the prepared DW (point (5) in Fig. 1(b)) as a function of an applied in-plane field and for current pulses applied parallel or perpendicular to the DW (inset of Fig. 1(a)). We fix the current pulse width to be 100 μs and the pulse current density to be j=± 6.4x10$^{10}$A/m². The action of the current is to reduce the purely field-induced switching field $H_c^*$ of the Hall cross to smaller absolute values $H_{dep}$. From this we can determine the current equivalent effective out-of-plane field $|\Delta H_z^{eff}|=|H_c^*|-|H_{dep}|$, which is a function of the applied in-plane field direction denoted by the angle ϕ as well as of the applied current density j.

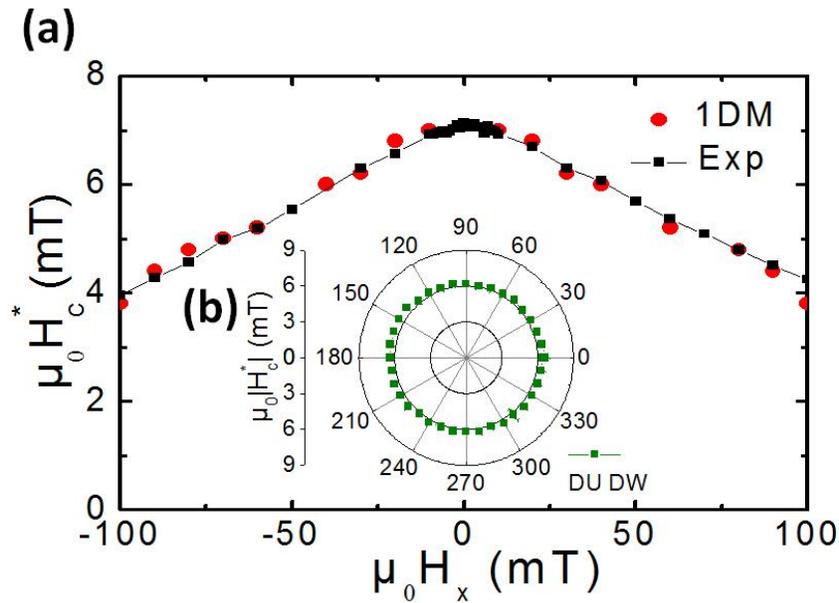

*Figure 2. (a) Out-of-plane depinning field dependence on the applied longitudinal in-plane field amplitude for a DU DW. The black squares indicate the experimental results, which are in excellent agreement with the 1D model[8] (red dots). (b) The inset shows the angular dependence of depinning field for a fixed in-plane field amplitude of 40 mT, showing an isotropic behavior.*

We first determine the pure field-induced depinning of a DU DW by measuring the out-of-plane depinning field $H_c^*$ as a function of an applied in-plane field in the ±x-direction (Fig. 2 (a)). At zero in-plane field the switching field is around 7 mT. It decreases with an applied in-plane field with a slope of $|H_c^*/H_x| \approx 3/100$. Next, we performed an angular scan in which the depinning field is measured for all directions (0 to 360 degrees) of the in-plane field. An angular scan of the switching field at a fixed in-plane field amplitude of 40 mT (Fig. 2 (b)) shows that the field-induced depinning is completely isotropic in the polar plane, which indicates that its mechanism does not depend on the internal spin-structure of the DW. This is in line with the results of calculations using the 1D model where the in-plane field changes the domain wall spin structure leading to a larger efficiency of the out-of-plane field thus reducing the depinning field (Fig. 2(a)).

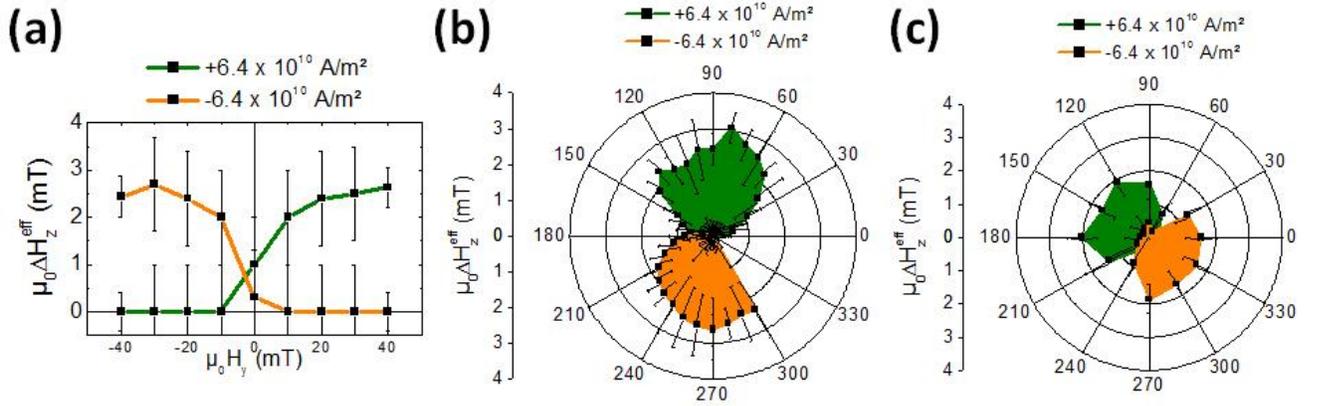

*Figure 3. Effective fields generated by the current $\Delta H_z^{eff}$ for DU DWs as a function of applied in-plane field and different current directions. (a) Current is applied in the ± y-direction, parallel or anti-parallel to the DW and parallel or anti-parallel to the in-plane field $H_y$. (b) Angular dependence of $\Delta H_z^{eff}$ at a fixed in-plane field amplitude of 40 mT and a current applied in the ± y direction. (c) As in (b), but for current along the ± x direction. The plots show a strongly anisotropic behavior.*

Having established that the field-induced depinning is isotropic for in-plane fields along different directions we next study the effect of currents, as any anisotropic behavior must thus result from the current - effects.

We first probe the novel geometry with a perpendicular (x-direction) displacement of a DW when a current is applied parallel or anti-parallel to the DW (along the ±y-direction, geometry see inset in Fig. 1 (a)). In this case, we use a field along the ±y-direction, which means that at large fields the wall is a pure BW. In Fig. 3 (a), we see that the effective field $\Delta H_z^{eff}$ resulting from the current increases for the combination of a positive (negative) current density and a positive (negative) $H_y$-field and saturates at $|H_y| \approx 40$ mT. The resulting maximum SOT-efficiency is $\chi \approx (3.9$ mT$) / (10^{11}$ A/m²$)$. The saturation at 40 mT shows that for this field value the DW has pure BW structure due to the applied field, while at zero in-plane field the DMI generates a Néel component[8]. This increase in the efficiency with increasing BW structure can be attributed to a dominating damping-like torque, which is expected from the SHE[7]. To test this, we performed an angular scan with a fixed in-plane field amplitude of 40 mT and a current applied parallel or anti-parallel to the DW. Fig. 3 (b) shows the experimental results of the angular dependence of the depinning efficiency for a DU DW. Such dependence is strongly anisotropic and depends on the applied in-plane-field direction, in contrast to the purely field-induced depinning case. The angular dependence with a maximum at 90° and 270° is compatible with a damping-like SOT that could for instance result from the SHE.

To check if additional torques beyond SOTs as allowed by symmetry[22] are present, we next investigated the torques acting for a current flowing perpendicular to the DW (j along x-axis).

Fig. 3 (c) shows the experimental results of an angular scan of the depinning field for a current injected perpendicular to the wall. We find that the maximum efficiency is similar to the case for the current parallel to the DW. However, in this case, we find the highest current-driven mobility for DWs of NW-type, which is consistent with a SHE torque as the dominant mechanism behind the current-induced motion[26]. Generally, we find (within the error bars) the same maximum SOT efficiencies for currents applied parallel and perpendicular to the DWs.

We next discuss these finding by analyzing the acting torques. Assuming a dominating damping-like torque the spin dynamics reads[7,10]:

$$\frac{\partial \boldsymbol{M}}{\partial t} = -\gamma \boldsymbol{M} \times \boldsymbol{H} + \frac{\alpha}{M_s} \boldsymbol{M} \times \frac{\partial \boldsymbol{M}}{\partial t} - (\boldsymbol{u} \cdot \boldsymbol{\nabla})\boldsymbol{M} + \frac{\beta}{M_s} \boldsymbol{M} \times (\boldsymbol{u} \cdot \boldsymbol{\nabla})\boldsymbol{M} + \frac{\alpha_{SHE}}{M_s} \boldsymbol{M} \times (\boldsymbol{\sigma} \times \boldsymbol{M})$$

where $M_s$ is the saturation magnetization, $\alpha$ is the Gilbert damping parameter, $\beta$ is the non-adiabaticity parameter, $\gamma$ is the gyromagnetic ratio, **u** is proportional to the charge current density **j**, $\sigma$ denotes the SHE spin direction and $\alpha_{SHE}$ is a parameter determining the strength of the SHE resulting in the damping-like torque. For negligible STT[6], the SOT-model predicts that the torque acting on a NW for a current perpendicular to the wall and the torque acting on a BW for a current parallel to the wall is identical. And this holds even for combinations where the current is at an angle with the wall internal magnetization-direction. Therefore, our finding of similar torques for current parallel and perpendicular to the wall support the damping-like torque mechanism for the SOTs that lead to the DW depinning. Within the experimental uncertainty, we do not observe any other dominating torques such as those predicted theoretically by Hals and Brataas[22] by symmetry considerations in our Ta\CoFeB\MgO\Ta multilayer.

Finally, we note that the angle of the maximum efficiency is not always exactly along the expected direction. For example in Fig. 3 (c), one expects for a perfectly straight DW a maximum at 0° and 180. However, we see that the centers of the effective field areas are rotated by $\approx$ -30°. This can be explained by the fact that the DWs are pinned at certain pinning sites randomly distributed in the Hall cross. The details of the local pinning spot and its magnetic properties can result in a weak deformation of the DW structure, which may lead to non-collinear dependences with respect to the current direction. Furthermore micro-magnetic simulations indicate that the DW bends under the influence of fields and currents before the full depinning event and that the internal magnetization of the DW can be influenced by local variations of the geometry and magnetic properties, so that the 1D-model is not sufficient anymore to describe the depinning process as visible by our results.

However, using our full angular scans, we find that the maximum efficiency is always similar within the error bars, allowing us to conclude that the damping-like torque can explain the acting torques for all cases. To test this further, the exact geometry of the wall, the pinning position and the current flow distribution need to be known, which goes beyond the scope of this manuscript. However, by analyzing the full angular scans of the depinning fields, we can rule out novel dominating torques, which lead to vastly different efficiencies for currents parallel or perpendicular to the wall.

In summary, we studied the field- and current induced depinning of a DW in a Hall cross geometry. While the purely field-induced depinning shows no angular dependence with respect to the direction of the in-plane field, the effect of the current depends crucially on the internal DW structure, which we manipulate by an external magnetic in-plane field. We find a maximum of the SOT-efficiency for current flowing parallel to the DW for an in-plane field that leads to a BW, which is in line with an interpretation based on a damping-like SOT. By comparing the cases of current flowing parallel to the DW and perpendicularly to the DW, we find that the maximum efficiency is similar for these cases and no large new torques arise for currents parallel to the wall. This is in line with an explanation based on the damping-like torque with no large further torques that are allowed by symmetry and with the negligible STT in our system[6]. Finally, we find a varying angular dependence of the maximum depinning efficiency for different DWs and pinning levels. This emphasizes the importance of full angular scans and not just measurements for two field directions and shows the sensitivity of the SOT to the precise DW structure and pinning sites. To analyze the data further and identify possibly small new torques that we cannot exclude with the current measurements, a combined experiment together with spin-sensitive imaging techniques to determine the wall shape and pinning site is necessary, which is beyond the scope of this work but could lead to the discovery of new torque terms.

Finally, concerning possible technological applications, our demonstration of DW depinning in a Hall cross geometry by a current flowing parallel to a DW allows for new device geometries, which can be of interest for industrial applications as it opens a previously unexplored avenue to switching by DW motion.

## ACKNOWLEDGMENTS

We acknowledge support by the Graduate School of Excellence "Materials Science in Mainz"(MAINZ) GSC 266; the EU (IFOX, Project No. NMP3-LA-2012 246102; MASPIC, Project No. ERC-2007-StG 208162; WALL, Project No. FP7-PEOPLE-2013-ITN 608031; MAGWIRE, Project No. FP7-ICT-2009-5), and the Research Center of Innovative and Emerging Materials at Johannes Gutenberg University (CINEMA). E.M. acknowledges the support by Project No. MAT2011-28532-C03-01 from the Spanish government, and Project No. SA163A12 and Project No. SA282U14 from Junta de Castilla y Leon. Finally, we also thank the Singulus Technologies AG for support with the materials stacks preparation.